\documentclass[num-refs]{nbdt-article}

\usepackage{makecell}

\usepackage{siunitx}
\papertype{Original Article}

\title{Expressive architectures enhance interpretability of dynamics-based neural population models}

\author[1, 2]{Andrew R. Sedler}
\author[2]{Christopher Versteeg}
\author[1, 2]{Chethan Pandarinath}

\affil[1]{Center for Machine Learning, Georgia Institute of Technology, Atlanta, GA, USA}
\affil[2]{Wallace H. Coulter Department of Biomedical Engineering, Emory University and Georgia Institute of Technology, Atlanta, GA, USA}

\corraddress{Chethan Pandarinath, Ph.D., Wallace H. Coulter Department of Biomedical Engineering, Emory University and Georgia Institute of Technology, Atlanta, GA, USA}
\corremail{chethan@gatech.edu}

\fundinginfo{This work was supported by NSF Graduate Research Fellowship DGE-2039655 (ARS), the Emory Neuromodulation and Technology Innovation Center (ENTICe), NSF NCS 1835364, DARPA PA-18-02-04-INI-FP-021, NIH Eunice Kennedy Shriver NICHD K12HD073945, NIH-NINDS/OD DP2NS127291, NIH BRAIN/NIDA RF1 DA055667, the Alfred P. Sloan Foundation, and the Burroughs Wellcome Fund (CP), and the Simons Foundation as part of the Simons-Emory International Consortium on Motor Control (CP, CV)}

\runningauthor{Sedler, Versteeg, Pandarinath}

\begin{document}

\maketitle

\begin{abstract}

Artificial neural networks that can recover latent dynamics from recorded neural activity may provide a powerful avenue for identifying and interpreting the dynamical motifs underlying biological computation. Given that neural variance alone does not uniquely determine a latent dynamical system, interpretable architectures should prioritize accurate and low-dimensional latent dynamics. In this work, we evaluated the performance of sequential autoencoders (SAEs) in recovering latent chaotic attractors from simulated neural datasets. We found that SAEs with widely-used recurrent neural network (RNN)-based dynamics were unable to infer accurate firing rates at the true latent state dimensionality, and that larger RNNs relied upon dynamical features not present in the data. On the other hand, SAEs with neural ordinary differential equation (NODE)-based dynamics inferred accurate rates at the true latent state dimensionality, while also recovering latent trajectories and fixed point structure. Ablations reveal that this is mainly because NODEs (1) allow use of higher-capacity multi-layer perceptrons (MLPs) to model the vector field and (2) predict the derivative rather than the next state. Decoupling the capacity of the dynamics model from its latent dimensionality enables NODEs to learn the requisite low-D dynamics where RNN cells fail. Additionally, the fact that the NODE predicts derivatives imposes a useful autoregressive prior on the latent states. The suboptimal interpretability of widely-used RNN-based dynamics may motivate substitution for alternative architectures, such as NODE, that enable learning of accurate dynamics in low-dimensional latent spaces.

Code:~\href{https://github.com/snel-repo/expressive-latent-dynamics-paper}{\texttt{github.com/snel-repo/expressive-latent-dynamics-paper}}


\end{abstract}

\section{Introduction}
Advances in recording technologies over the past decade have enabled simultaneous monitoring of hundreds to thousands of neurons, providing a qualitatively different picture of population activity than was previously available \cite{stevenson_how_2011, juavinett2019chronically, jun_fully_2017}. These high-dimensional recordings often display \textit{latent dynamical structure}, which includes both co-activation across neurons and coordinated evolution over time. In an emerging framework, termed \textit{computation through dynamics}, latent dynamical structure is hypothesized as a key mechanism for implementing biological computations \cite{vyas_computation_2020, shenoy_measurement_2021, khona2022attractor}. According to this framework, a neural population constitutes a dynamical system that performs computations through its temporal evolution. Such dynamics have been implicated in computations underlying movement generation, decision making, and working memory \cite{vyas_computation_2020, shenoy_cortical_2013, jazayeri_interpreting_2021}.

Based on these findings, a promising avenue toward understanding computation in the brain is to build models that can distill latent dynamical structure from recordings of neural activity. Models that accurately capture latent dynamics could be dissected via flow field visualization or fixed-point linearization to identify dynamical motifs that implement units of computation \cite{sussillo_opening_2013, driscoll2022flexible}. Toward this goal, a variety of methods have been developed to train latent dynamical models that reproduce recorded neural population activity \cite{linderman_bayesian_2017, sussillo_lfads_2016, schimel_ilqr-vae_2021, kim2021inferring, hurwitz2021building, duncker_dynamics_2021}. To date these methods have been highly successful in de-noising activity, for example producing representations of neural activity that correspond with subjects’ behaviors on a moment-by-moment basis and millisecond timescale \cite{pandarinath_inferring_2018, keshtkaran2022large, zhu2022deep}. Yet the degree to which the trained models enable interpretation of the biological circuit’s dynamics remain unclear.

To enable interpretation of biological computation, models must recover \textit{accurate} and \textit{parsimonious} representations of dynamics. Neural population models are often trained to reconstruct observed activity, with the assumption that better reconstruction performance corresponds to higher dynamical accuracy. However, because many different dynamical mechanisms could produce the same activity patterns at their output \cite{kao2019considerations}, there is no guarantee that a model with high reconstruction performance will have dynamics that match the system being modeled, opening the door to misinterpretation. Thus, it is critical to test whether different architecture choices lead to more accurate models of underlying dynamics. Additionally, empirical evidence suggests that the latent dynamics of neural populations are typically much lower dimensional than the number of neurons we observe \cite{yu_gaussian-process_2008}. Thus, models of latent dynamics should ideally match the dimensionality of the system being modeled (i.e., be parsimonious) to promote ease of interpretation.

One recent approach, Poisson Latent Neural Differential Equations (PLNDE \cite{kim2021inferring}), adapted neural ordinary differential equations (NODEs \cite{chen_neural_2019}) for the statistics of neural spiking data. PLNDE demonstrated impressive performance in recovering low-dimensional phase portraits and fixed points of nonlinear dynamical systems with few trials and low firing rates. While this work hinted that NODE-based models may perform better than recurrent neural network (RNN)-based models at lower latent state dimensionalities, it remains unclear how the NODE compares directly to the RNN, and what features of NODEs are responsible for the differences.

A salient difference between RNNs and NODEs is that the capacity of an RNN is tied to its latent dimensionality, while NODEs can define vector fields using networks of arbitrarily high capacity. This might enable NODE-based models to more easily learn dynamics in low-dimensional spaces, which would explain their performance advantage in that regime. Another distinction is that NODEs form predictions by adding inferred derivatives to the previous latent state via \textit{skip connections}, while RNNs infer the next state directly. Gated RNN variants have some capacity to learn gates that incorporate the previous state into their prediction, but they typically use a learned convex combination of a previous state and a predicted state. In NODEs, skip connections impose the inductive bias that a given state should be near its previous state. This should make optimization easier when the goal is to reconstruct a dynamical system.

In this work, we evaluate the performance of RNN-based and NODE-based SAEs in recovering latent chaotic attractors from simulated spiking datasets. We quantify the accuracy of inferred firing rates using the data log-likelihood and the true rate $R^2$. We also introduce a metric, state $R^2$, which measures the fraction of inferred latent state variance explained by an affine transformation of the true latent states. Despite their success in reconstructing neural activity patterns \cite{pandarinath_inferring_2018, keshtkaran2022large, schimel_ilqr-vae_2021}, we find that RNN-based SAEs require many more latent dimensions than the synthetic systems they are attempting to model. Moreover, we find that the dynamics learned by the RNNs are a poor match to the synthetic systems, in that a large fraction of the models’ variance reflects activity not seen in the synthetic system. Finally, we find that the behavior of RNNs linearized around their fixed points is qualitatively different from that of the true systems.

In contrast, we find that NODE-based SAEs learn dynamics in the true latent dimensionality with minimal superfluous dynamics, while capturing the behavior of the true system around its fixed points. Ablation experiments indicate that both more expressive update layers and skip connections are responsible for the NODE's ability to more effectively learn latent dynamics. The suboptimal interpretability of widely-used RNN-based dynamics may motivate substitution for NODE-like architectures that enable learning of accurate dynamics in low-dimensional latent spaces. Future models could leverage these characteristics to learn simpler, more intepretable latent dynamics that elucidate the computational mechanisms implemented by real neural populations.

\section{Related Work}
Converging theoretical and experimental evidence suggests that the time-course of neural activity can be modeled as a dynamical system whose state variables are not individual neurons, but instead a lower-dimensional set of latent variables \cite{duncker_dynamics_2021, pei_neural_2022}. In this paper we restrict analyses to systems that are well-modeled as autonomous dynamical systems, which can be described in continuous time as:

\begin{gather}
    \frac{d\mathbf{z}}{dt} = f(\mathbf{z}) \\
    \mathbf{x} \sim \text{Poisson}(\exp(g(\mathbf{z})))
\end{gather}

Where $z\in\mathbb{R}^D$ is the latent state, $x\in\mathbb{R}^N$ is the observed neural population activity, $f$ defines the latent dynamics that determine the time-evolution of the system, and $g$ defines the mapping from latent variables to neural activity. As neurons cannot have negative firing rates, we make outputs from $g$ positive using $\exp$ and follow with Poisson sampling to determine the number of times each neuron spikes in each time bin.

Early efforts to explicitly model the latent dynamics of neural populations imposed strict limitations on dynamical complexity by representing $f$ using a linear dynamical system (LDS) and $g$ using a linear readout for tractability and ease of interpretability \cite{macke2011empirical}. Subsequent approaches allowed more flexibility in $g$, enabling models to explain more variance in neural data while maintaining linear dynamics \cite{gao_linear_2016}. Still other approaches have used switching linear dynamical systems (SLDS) to approximate nonlinear dynamics in $f$ \cite{linderman_bayesian_2017, petreska2011dynamical}.

The field of dynamical system identification has had great success in recovering governing equations from time series measurements, particularly in the physics domain \cite{brunton2016discovering, champion2019data, bakarji2022discovering}. However, existing approaches rely heavily on sparse, symbolic regression from states to derivatives \cite{brunton2016discovering}, which poses a number of problems for spiking datasets. First, discrete counts and their derivatives are not suitable for regression and naive smoothing approaches are unlikely to preserve dynamically relevant features. Second, it is unclear whether the dynamics of neural populations can or should be described by a sparse selection from a library of nonlinear terms.

With the resurgence of deep learning over the past decade, a powerful class of methods has emerged that use RNNs to approximate $f$ \cite{pandarinath_inferring_2018, keshtkaran2022large, andalman2019neuronal, schimel_ilqr-vae_2021}. In head-to-head comparisons, RNN-based methods replicate neural activity patterns with substantially higher accuracy than LDSs on datasets from a variety of brain areas and behaviors, suggesting that linear dynamics may not adequately model the dynamics of neural systems \cite{pei_neural_2022}. A recent model combined the RNN and SLDS architectures to improve the RNN's interpretability, but it is unclear how this model compares to standard RNNs in recovering latent dynamics \cite{smith_reverse_2021}. 

Early characterization of the dynamics of RNN-based models for interpretability began with fixed-point linearizations of networks trained to perform simple tasks, such as 3-bit memory and pattern generation \cite{sussillo_opening_2013}. Subsequent work has used similar techniques to elucidate the computational dynamics underlying text classification and to illustrate commonalities among different types of RNNs trained to perform the same task \cite{maheswaranathan_reverse_2019, maheswaranathan_universality_2019}.

It is worth noting that some powerful, approaches for interpreting neural activity consider latent embeddings for each time step individually and are agnostic to temporal dynamics. Some such models rely on variational penalties \cite{higgins2017beta}, identifiability constraints \cite{moran_identifiable_2023}, and/or behavioral information \cite{schneider2022cebra} to impose structure on the latent space. While these models are a viable alternative for discovering structure in neural activity, they do not provide any direct estimate of neural dynamics and therefore don't permit estimates of dynamical features, such as fixed points, that may provide insight into the computations performed by neural populations \cite{vyas_computation_2020}.

The goal of this work is to directly compare RNN-based and NODE-based SAEs to clearly present the benefits of more interpretable architectures for neural population models.

\section{Methods}
\subsection{Datasets}
To determine whether trained models accurately recapitulate latent dynamical rules from observed population activity, we require reference datasets where the ground truth dynamics are known. Accordingly, we set up synthetic test cases that mimic empirical properties of neural systems, i.e., low-D latent dynamics that are observed via noisy spiking activity of neural populations. We created each synthetic dataset based on a long simulated trajectory of a chaotic dynamical system ($D=3\;\text{or}\;10$; see Supplementary Table 1). To generate firing rates for $N=10\;\text{or}\;100$ neurons, the trajectory was first projected through a $D\times N$ random matrix with elements sampled from $U[-0.5, 0.5]$. The resulting data were standardized to zero-mean and unit variance and passed through an exponential activation function, resulting in $N$ time-varying firing rates. To simulate spiking activity, firing rates were sampled as inhomogenous Poisson processes. From this spiking data, 1600 segments of length $T=70$ were randomly split into training ($1280\times T\times N$) and validation ($320\times T\times N$) datasets.

\subsection{Model Architecture}
All models used in this work were variations of a sequential autoencoder (SAE). Briefly, the binned spike counts, $\mathbf{X}\in\mathbb{R}^{T\times N}$, were fed into a 64-unit bidirectional gated recurrent unit (BiGRU) layer, whose final hidden states were linearly projected to an initial condition, $\hat{\mathbf{z}}_0$.

\begin{gather}
    \mathbf{h}_T = \textbf{BiGRU}(\mathbf{X}) \\
    \hat{\mathbf{z}}_0 = \mathbf{W}_E \mathbf{h}_T + \mathbf{b}_E
\end{gather}

The initial condition was unrolled by the dynamics model ($f$, GRU or NODE) to compute the time-varying latent states, $\hat{\mathbf{z}}_t$, which were projected to inferred rates, $\hat{\mathbf{y}}_t$, by a linear-exponential readout:

\begin{gather}
    \hat{\mathbf{z}}_t = \textbf{GRU}(\hat{\mathbf{z}}_{t-1})
        \quad\quad\text{OR}\quad\quad
    \hat{\mathbf{z}}_t = \hat{\mathbf{z}}_{t-1} + \int_{t-1}^t\textbf{NODE}(\hat{\mathbf{z}}(t))dt \label{eq:dyn_fwd}\\
    \hat{\mathbf{y}}_t = \exp(\mathbf{W}_R \hat{\mathbf{z}}_t + \mathbf{b}_R)
\end{gather}

We use standard GRU cells throughout the model (see Section \ref{supp:gru}). The NODE consisted of an MLP with one hidden layer of 128 $\tanh$ units. For the GRU, it is important to note that both model capacity and dimensionality of the state space are determined by the number of hidden units, and dynamics are determined by the recurrent connectivity. On the other hand, the NODE separates state space dimensionality from model capacity and explicitly models dynamics by approximating a vector field using an MLP. Additionally, note that the NODE explicitly incorporates the previous state into each prediction.

\begin{equation} \label{eq:node_mlp}
    \textbf{NODE}(\hat{\mathbf{z}}) = \mathbf{W}_2 \tanh(\mathbf{W}_1 \hat{\mathbf{z}} + \mathbf{b}_1) + \mathbf{b}_2
\end{equation}

Networks were trained to maximize the Poisson log-likelihood of the observed data, $\mathbf{X}=[\mathbf{x}_1 \dots \mathbf{x}_T]^T$, given the inferred firing rates, $\hat{\mathbf{Y}}=[\hat{\mathbf{y}}_1 \dots \hat{\mathbf{y}}_T]^T$. 

\begin{equation}
    L_\theta(\hat{\mathbf{Y}}) = - \frac{1}{TN}\sum_{t=1}^T\sum_{n=1}^N\log\text{Poisson}(x_{t,n} | \hat{y}_{t,n})
\end{equation}

While GRU outputs were computed with a fixed time step, NODE trajectories were computed using a Runge--Kutta solver \cite{tsitouras2011runge} (except for Mac Arthur and ablation experiments, which were discretized at the bin width) and were trained incrementally to avoid dynamical collapse to the origin. Rather than compute loss on the whole trajectory, we added groups of new time steps at regular intervals, up to the max of $T$ steps (Supplementary Table \ref{supp:hp_table}). We tuned batch size, learning rate, and weight decay to find models that performed well on validation data across a range of latent sizes. All networks were trained via backpropagation through time. Further training details are given in the Supplementary Material.

\subsection{Metrics}
All metrics were evaluated on validation data. Reconstruction performance was evaluated by two key metrics. The first, spike negative log-likelihood (NLL), was the same as the Poisson NLL used during model training. The second, rate $R^2$, was the coefficient of determination between true and inferred firing rates, averaged across neurons. While these metrics capture similar information, we use both because spike NLL is the only metric available for real datasets, but rate $R^2$ captures the model’s ability to denoise the data.

While reconstruction performance evaluates how well the inferred latent states explain the variance in the neural data, it is blind to variance in the inferred latent states that is not explained by the true latent states. We introduce a new metric, state $R^2$, that provides a complementary view of the accuracy of the inferred latent states. State $R^2$ measures the fraction of inferred latent state variance explained by an affine transformation of the true latent states. Taken together, rate $R^2$ and state $R^2$ indicate whether models are inferring the latent features of the true system, without inventing superfluous features that could hinder interpretability.

\section{Results}
\subsection{Neural ODEs learn more accurate and parsimonious representations of dynamics}
\begin{figure}[ht]
  \centering
  \includegraphics[width=\textwidth]{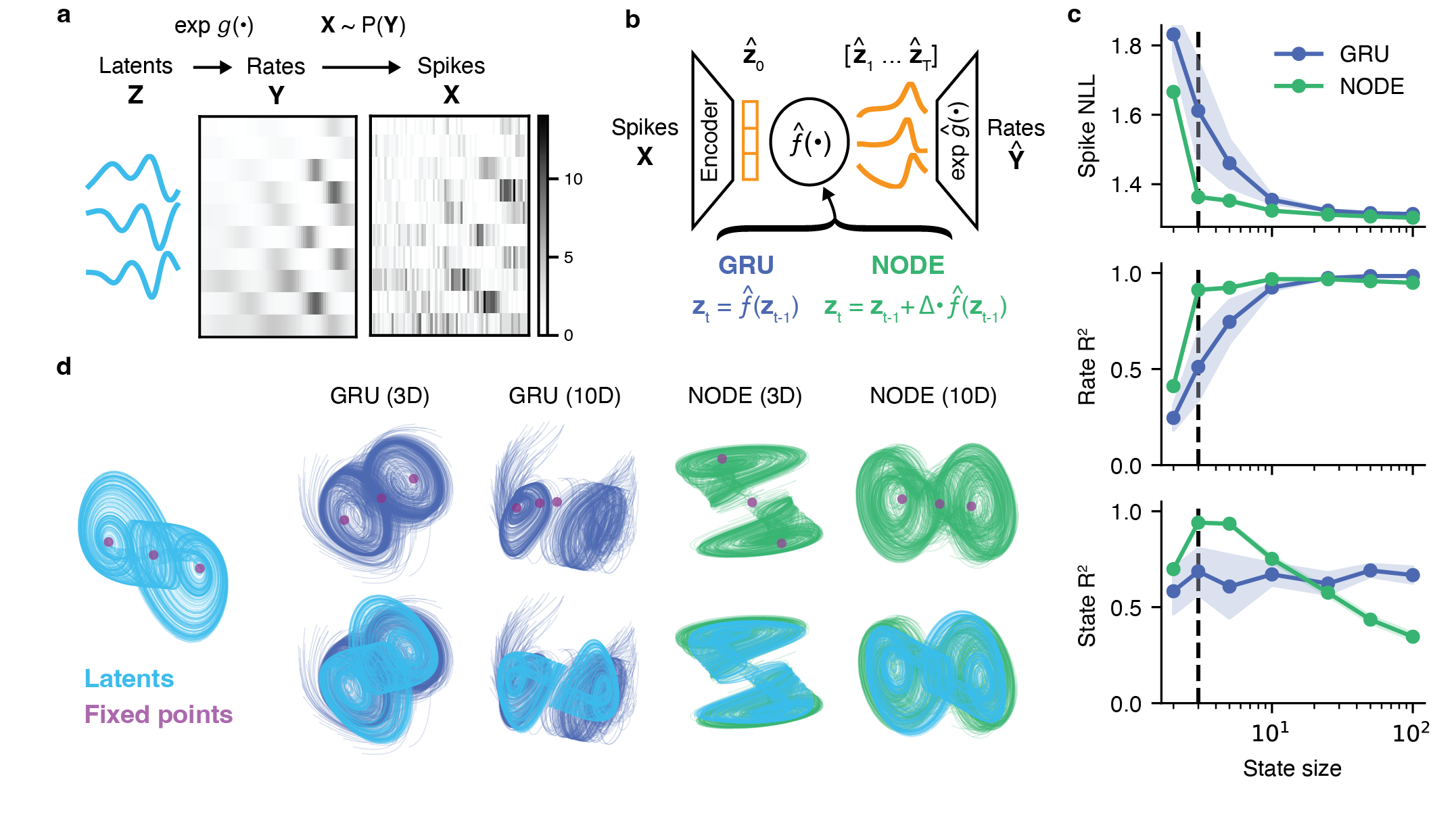}
  \caption{Reconstruction of Arneodo system using GRUs and NODEs. (a) The data generation pipeline for a single sample consisted of projecting trajectories through a random linear projection, $g$, followed by exponentiation and Poisson sampling. (b) A schematic showing the architecture of GRU-based and NODE-based SAEs used in these experiments. (c) Performance metrics for models of differing sizes. Traces are means and shaded regions are standard deviations across five different initializations. Vertical line indicates the true latent dimensionality. (d) Latent trajectories inferred by each model (or top-3 PCs), overlaid with learned fixed points (top) and with the latent projection used to compute state $R^2$ (bottom).
  }
  \label{fig:arneodo1}
\end{figure}

We trained RNN-based and NODE-based SAEs to model the latent dynamics underlying synthetic datasets, for which ground truth firing rates, latent states, and fixed points were known. We created a synthetic dataset by projecting an Arneodo system trajectory into an observation space and sampling binned spike counts (Fig. \ref{fig:arneodo1}a; see Methods). The Arneodo system is a mildly chaotic 3-D attractor based on a modified Lotka-Volterra ecosystem (see Supplementary Material). We chose this system because (1) it demonstrates relatively limited chaotic behavior \cite{arneodo_occurence_1980} and (2) regions around its fixed points are well-sampled by the latent trajectories, making them more easily identifiable.

We evaluated reconstruction performance as a function of state size for GRU-based and NODE-based autoencoders (Fig. \ref{fig:arneodo1}b). The reconstruction performance of both model types increased with state size, but the GRU consistently required higher dimensionality to achieve similar reconstruction performance to the NODE. For example, GRU-based models required a state size of 10 in order to match the performance of the 3-D NODE (Fig. \ref{fig:arneodo1}c, top and middle). This indicates that NODEs are better suited than GRUs for modeling dynamics in low-dimensional latent spaces.

While low-dimensional GRUs do not have sufficient capacity to model the data, it is possible that high-dimensional GRUs could recover the low-dimensional attractor embedded in a higher-dimensional state space. In this case, we would expect the true latent states to explain the variance in the inferred latent states. To quantify this phenomenon we introduce a new metric, state $R^2$ (see Methods), which evaluates the fraction of the variance of inferred latent states explained by an affine transformation of the true latent states. In combination with reconstruction performance, it allows us to determine which models are recovering the true latent trajectories without inventing dynamics that do not exist in the true system. 

Despite accurate reconstruction performance at high dimensionalities, GRUs exhibited poor state $R^2$ for all models, remaining relatively flat across state sizes and peaking at $R^2=0.69$ for the 50-D model. This implies that, even as larger GRUs exhibited high reconstruction performance, a substantial fraction of the GRU's variance remained inconsistent with the true dynamics. In contrast, the 3-D NODE scored highest by our state $R^2$ metric ($R^2=0.94$) in comparison to all tested alternatives, and increasing the latent size to 5-D caused only a minor decrease (Fig. \ref{fig:arneodo1}c, bottom). For these models, the variance in NODE states could be explained almost exclusively by the true latent states. The 2-D NODE had low state $R^2$ because it was not able to capture the true system, and larger NODEs saw gradual declines as models increasingly relied on inaccurate dynamics.

The superfluous and inaccurate dynamics of the GRU models were evident in visualizations of the inferred latent trajectories (Fig. \ref{fig:arneodo1}d). The presence of inaccurate dynamics despite high reconstruction performance provides a warning for interpreting the dynamics of these models in real neural datasets, where no ground truth latent system exists and state $R^2$ cannot be evaluated. In such cases, analysis of inaccurate dynamics could lead to erroneous conclusions about the computational mechanisms of the neural population.

\begin{figure}[ht]
  \centering
  \includegraphics[width=0.9\textwidth]{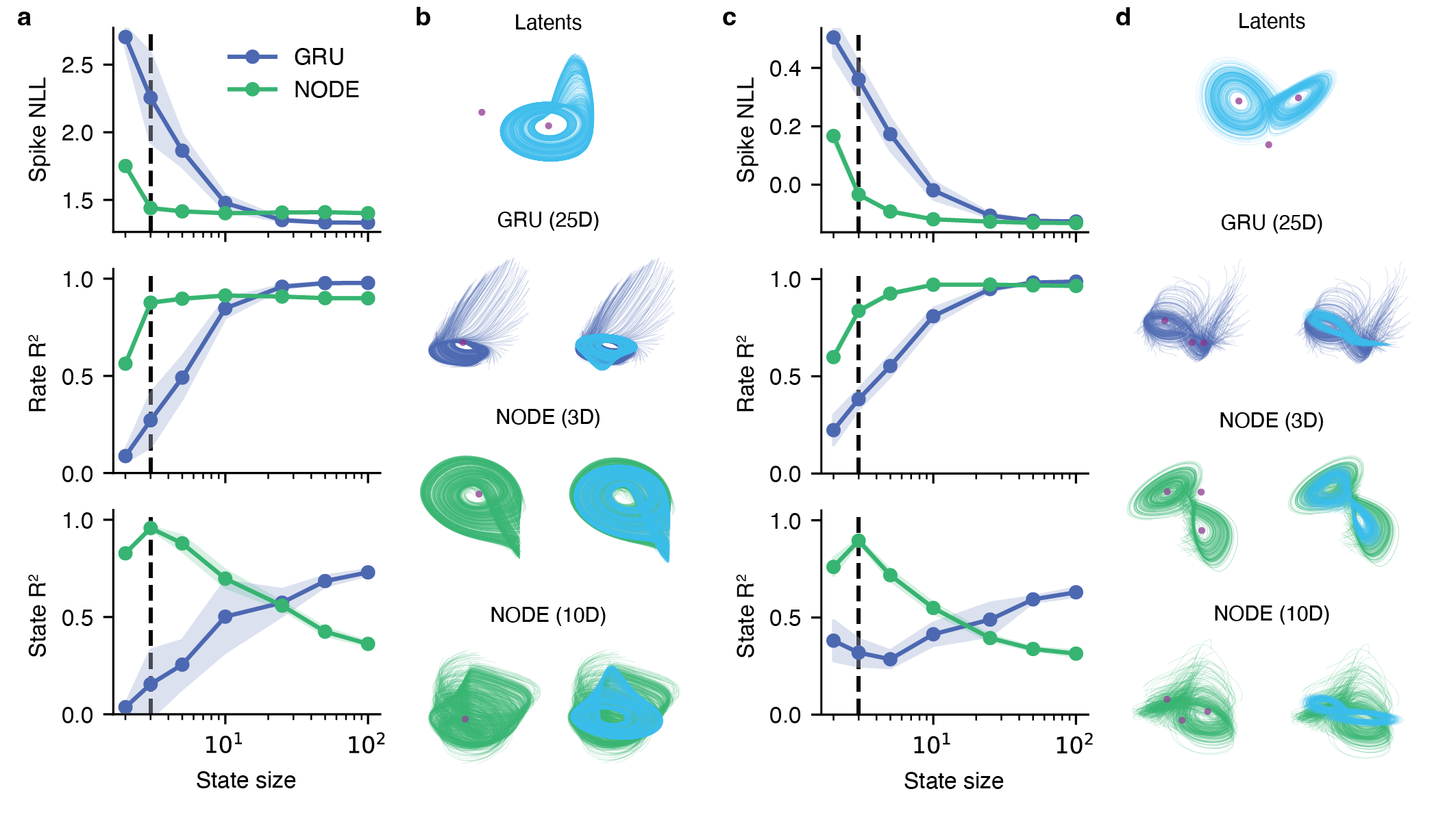}
  \caption{Reconstruction of R\"{o}ssler and Lorenz systems using GRUs and NODEs. (a) Performance metrics for models trained on the R\"{o}ssler dataset. Traces are means and shaded regions are standard deviations across five different initializations. (b) Latent trajectories inferred by each model for the R\"{o}ssler dataset (or top-3 PCs), overlaid with learned fixed points (left) and with the latent projection used to compute state $R^2$ (right). (c-d) Same as (a-b), for the Lorenz dataset.}
  \label{fig:rossler_lorenz}
\end{figure}

\subsection{Neural ODEs learn better representations of dynamics across systems}
We next applied the same SAE models and analysis to synthetic data from two additional dynamical systems, the R\"{o}ssler and Lorenz attractors, and found that NODE continued to achieve high reconstruction performance with fewer latent dimensions and fewer superfluous dynamical features than the GRU (Fig. \ref{fig:rossler_lorenz}). State $R^2$ was clearly the highest for the NODE-based models with the correct dimensionality (3-D) across all datasets. However, in real neuroscientific datasets, the "correct" dimensionality is unknown and state $R^2$ is unavailable, thus model selection would have to rely on spike NLL. For the R\"{o}ssler and Arneodo systems, the spike NLL of the NODE produced an elbow at the correct dimensionality (Figs. \ref{fig:rossler_lorenz}a, top and \ref{fig:arneodo1}c, top), which could serve as a guide for selecting the appropriate dimensionality. However, NODE models of the Lorenz system tended to overestimate latent dimensionality, as 5-D and 10-D models performed substantially better than 3-D (Fig. \ref{fig:rossler_lorenz}d, top).

For R\"{o}ssler and Lorenz systems, GRU state $R^2$ performance tended to increase with dimensionality up to a plateau, which likely reflected an increasing ability to fit the data. The fact that GRU required such high dimensionality to fit simple 3-D systems makes it unlikely to be an interpretable model of real neural systems. Notably, state $R^2$ on the Lorenz dataset plateaued and began to decline for models with states larger than 100 units (not shown).

\begin{figure}[ht]
    \centering
    \includegraphics[width=0.8\textwidth]{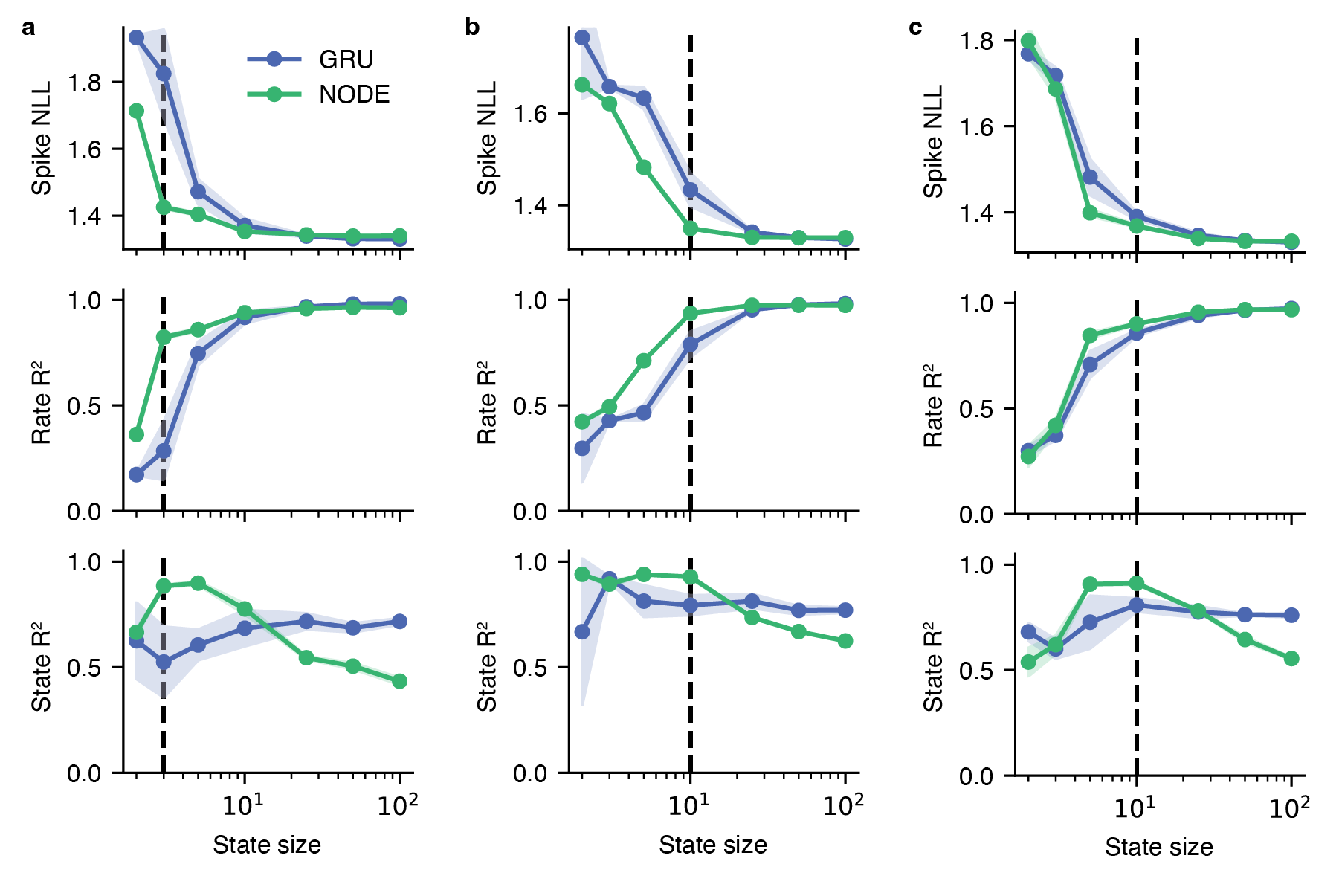}
    \caption{Reconstruction of latent systems using GRUs and NODEs for large datasets. (a) Performance metrics for models of differing sizes on a larger Arneodo dataset ($N=100$ neurons). Traces are means and shaded regions are standard deviations across five different initializations. Vertical line indicates the true latent dimensionality. (b) Same as (a) for a large Mackey-Glass dataset ($D=10$ latent dimensions, $N=100$ neurons). (c) Same as (a) for a large Mac Arthur dataset ($D=10$ latent dimensions, $N=100$ neurons).
    }
    \label{fig:100neurons}
\end{figure}

To confirm that these results generalized to datasets with more typical neuron counts and larger latent systems, we performed identical experiments with $N=100$ neurons for an Arneodo system (Fig. \ref{fig:100neurons}a), a Mackey-Glass system ($D=10$; Fig. \ref{fig:100neurons}b), and a Mac Arthur system ($D=10$; Fig. \ref{fig:100neurons}c). The Mackey-Glass system is a delay differential equation that represents a physiological circuit with time-delayed feedback \cite{glass1979pathological}, and the Mac Arthur system models a set of species competing for a set of renewing resources \cite{macarthur1969species, huisman1999biodiversity}. Under these conditions, NODE-based SAEs still required fewer dimensions to reconstruct neural activity and inferred more accurate latent trajectories than GRU-based SAEs. For the Mac Arthur dataset, we tried increasing the depth of the NODE MLP to two hidden layers and the encoder dimensionality to 128 and found that this slightly improved performance of the NODE relative to the GRU, without changing the latent state dimensionality (not shown).

\begin{figure}[ht]
  \centering
  \includegraphics[width=\textwidth]{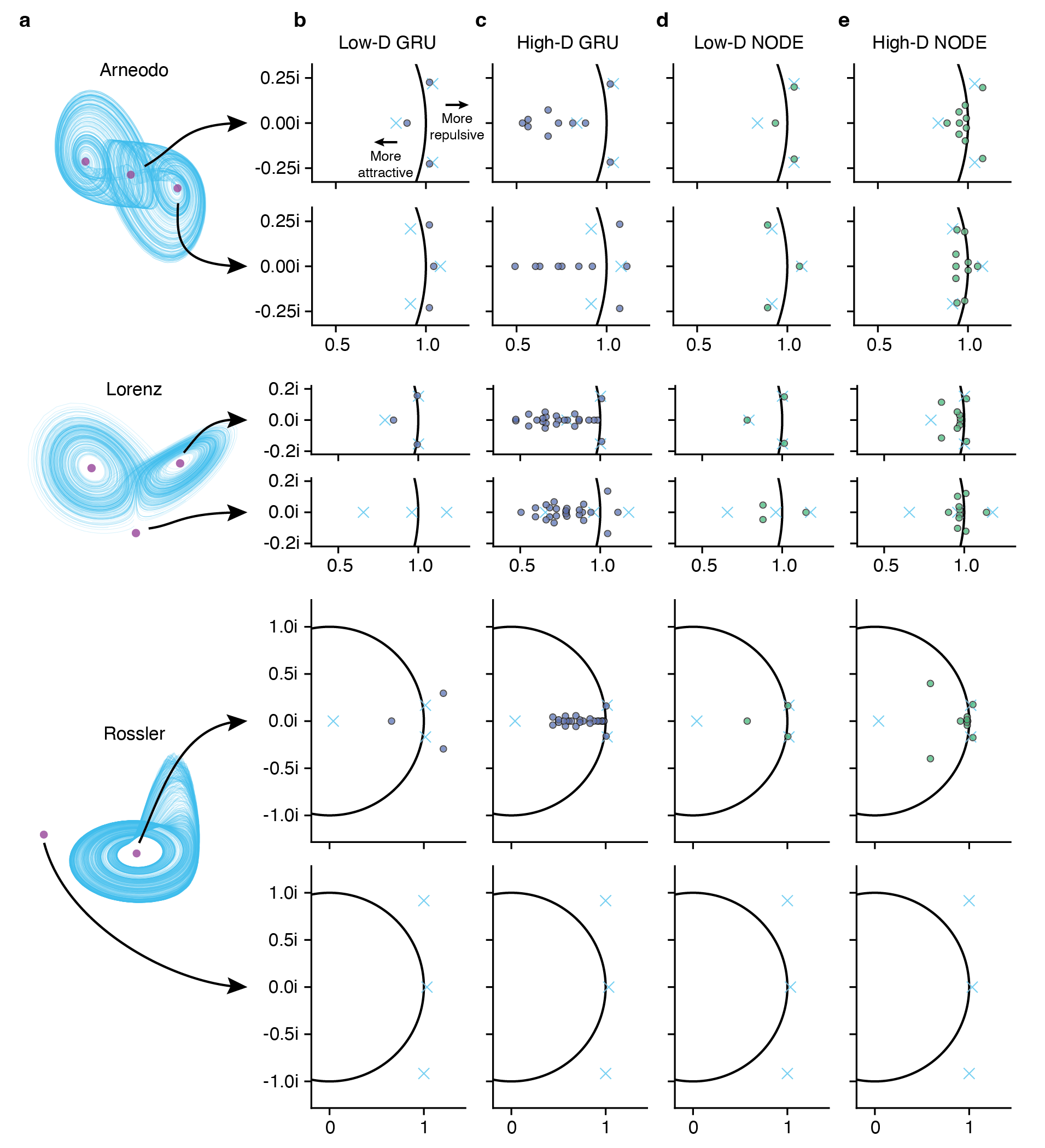}
  \caption{Recovery of fixed points and their properties using GRUs and NODEs. (a) True latent trajectories and fixed point locations for the three modeled systems. (b) Fixed points of trained networks were found by gradient descent on the norm of the learned vector field. Discrete-time Jacobians were computed for each fixed point of each network on each dataset. Jacobian eigenvalues are plotted here in the complex plane for the 3-D GRUs. Markers indicate the true (cyan $\times$) and estimated (colored circle) eigenvalues. (c) Same as (b) for the 10-D GRU (Arneodo) and 25-D GRUs (Lorenz and R\"{o}ssler). (d) Same as (b) for the 3-D NODEs. (e) Same as (b) for the 10-D NODEs.}
  \label{fig:all_fps}
\end{figure}

\subsection{Neural ODEs more effectively recover fixed point properties}
Going beyond comparing trajectories, the fixed points (FPs) of nonlinear dynamical systems are useful interpretability tools because they identify topologically important, locally linear regions where the system Jacobian can be decomposed to identify dynamical modes. The linearized dynamics around an FP indicate whether a nearby trajectory will diverge from ($||\lambda_i|| > 1$), converge to ($||\lambda_i|| < 1$), or oscillate around ($\Im(\lambda_i) > 0$) that FP along a given eigenvector, $\mathbf{v}_i$, providing a qualitative and succinct description of the system's dynamics in that region. By pairing learned FPs to their corresponding true FPs and comparing eigenvalues, we reveal more fine-grained information about whether the learned dynamics have similar behavior to the true system.

For each dataset, we selected representative low-dimensional and high-dimensional models with GRU-based and NODE-based dynamics. All of these models achieved high reconstruction performance, with the exception of the low-dimensional GRUs. For each model, we found a set of FPs by minimizing the norm of the vector field, as demonstrated in previous work \cite{sussillo_opening_2013, Golub2018}. This approach identified three FPs for Arneodo and Lorenz datasets and one FP for the R\"{o}ssler dataset in all models with high reconstruction performance. At each FP, we computed the discrete-time Jacobian of the corresponding dynamics model. Discrete-time Jacobians for the true systems were computed analytically. All Jacobians were decomposed to their eigenvalues, which were the subjects of analysis. Learned FPs were trivially mapped to the corresponding true FPs based on their eigenspectrum. For Arneodo and Lorenz datasets, we analyzed only one FP from each symmetric pair. We compare the properties of each learned FP to those of the corresponding true FP by overlaying plots of their eigenvalues in the complex plane (Fig. \ref{fig:all_fps}).

Models learned FPs analogous to those of the true systems with varying degrees of fidelity. The low-D NODE matched FPs and their qualitative behavior very well overall (Fig. \ref{fig:all_fps}d). On the other hand, the low-D GRU was not able to reconstruct the data or adequately recover FPs on any of the datasets (Fig. \ref{fig:all_fps}b). For example, the model learned lobe FPs with unstable oscillations ($||\lambda_i|| > 1$) instead of the stable oscillations of the true system ($||\lambda_i|| < 1$). The high-D GRU achieved good reconstruction performance, but in all cases developed FPs with many attracting dimensions ($||\lambda_i|| < 1$) (Fig. \ref{fig:all_fps}c). These additional attractive mechanisms allowed the initial conditions to be placed far away from the regions of state space frequented by the model, creating the trailing trajectories visible in the PCA projections (Fig. \ref{fig:arneodo1}d). A more accurate model would have mostly stable dimensions ($\lambda_i \approx 1$), which would prevent trailing trajectories and force initial conditions to be placed closer to their true locations in state space. The high-D NODE captured the core properties of the fixed points well in most cases, but tended to develop superfluous stable oscillations ($||\lambda_i|| < 1$ and $\Im(\lambda_i) > 0$) (Fig. \ref{fig:all_fps}e) which were also visible in the PCA projections (Fig. \ref{fig:arneodo1}d). All NODE and GRU models struggled to capture the central Lorenz FP, which had few nearby trajectories. Similarly, all models failed to capture a R\"{o}ssler FP that was distant from the system's trajectories. We believe this to be a limitation of the dataset, where the dynamics near some fixed points are underdetermined for the given trajectories, and highlights the need for more comprehensive sampling of latent state spaces for robust fixed point recovery.

% Clear the page so Fig. 3 stays near its text section
\clearpage

\subsection{Ablation of discrete time NODE-based SAE}
Several differences distinguish the NODE from the GRU trained in the preceeding sections, so it is not immediately obvious which of these is most responsible for performance gains. At a high level, these include (1) a continuous-time vs. discrete-time forward pass, (2) an incremental fitting process vs. training on the entire sequence, (3) combining the network output with the previous state via a skip connection vs. predicting a convex combination of a previous state and an updated state, and (4) use of an MLP vs. a linear layer followed by $\tanh$ to parameterize the update.

We have not found the NODE's continuous-time forward pass to be particularly important and for the sake of this comparison we discretize the NODE at the bin width. In the  rest of this section, we elaborate on and investigate the remaining differences.

Omitting inputs and switching to more standard RNN notation for clarity, the equations for the GRU are:

\begin{align}
    \mathbf{r}_t &= \sigma(\mathbf{W}_r \mathbf{h}_{(t-1)} + \mathbf{b}_r) \label{eq:gru_fwd_first} \\
    \mathbf{z}_t &= \sigma(\mathbf{W}_z \mathbf{h}_{(t-1)} + \mathbf{b}_z) \\
    \mathbf{n}_t &= \tanh(\mathbf{r}_t * (\mathbf{W}_n \mathbf{h}_{(t-1)} + \mathbf{b}_n)) \\
    \mathbf{h}_t &= \mathbf{z}_t * \mathbf{h}_{(t-1)} + (1 - \mathbf{z}_t) * \mathbf{n}_t \label{eq:gru_fwd_last}
\end{align}

Where $\mathbf{h}_t$ is the hidden state, $\mathbf{h}_{(t-1)}$ is the previous hidden state, $\mathbf{r}_t$ is the reset gate, $\mathbf{z}_t$ is the update gate, $\mathbf{n}_t$ is the candidate state, and $\sigma$ is the sigmoid function. This architecture is suboptimal for approximating a dynamical system for several reasons. First, without inputs, the reset gate $\mathbf{r}_t$ only serves to block the flow of information from $\mathbf{h}_{(t-1)}$ to $\mathbf{n}_t$. This is undesirable because $\mathbf{n}_t$ is the only mechanism the GRU has for updating the state. Second, there is no clear reason for the update gate $\mathbf{z}_t$ to mediate a convex combination of $\mathbf{h}_{(t-1)}$ and the candidate, $\mathbf{n}_t$, since intuitively $\mathbf{h}_t$ should generally be close to $\mathbf{h}_{(t-1)}$. Notably, when $\mathbf{r}_t=1$ and $\mathbf{z}_t=0$, the GRU simplifies to the vanilla RNN:

\begin{equation}
    \mathbf{h}_t = \tanh(\mathbf{W}_n \mathbf{h}_{(t-1)}+ \mathbf{b}_n)
\end{equation}

Rather than try to predict $\mathbf{h}_t$ \textit{de novo}, as in the vanilla RNN, or as a convex combination of $\mathbf{h}_{(t-1)}$ and a candidate state $\mathbf{n}_t$, the NODE allows a direct skip connection to $\mathbf{h}_{(t-1)}$. This builds in prior knowledge that $\mathbf{h}_t$ should be close to $\mathbf{h}_{(t-1)}$. Additionally, rather than assuming the update can be predicted using a linear layer followed by $\tanh$, the NODE allows the derivative to be computed by any network (we use a 2-layer MLP). Accordingly, the equation for the discretized NODE ($\Delta=0.1$) is:

\begin{equation}
    \mathbf{h}_t = \mathbf{h}_{(t-1)} + \Delta * \mathbf{W}_2 \tanh(\mathbf{W}_1 \mathbf{h_{(t-1)}} + \mathbf{b}_1) + \mathbf{b}_2
\end{equation}

\begin{figure}[ht]
  \centering
  \includegraphics[width=2in]{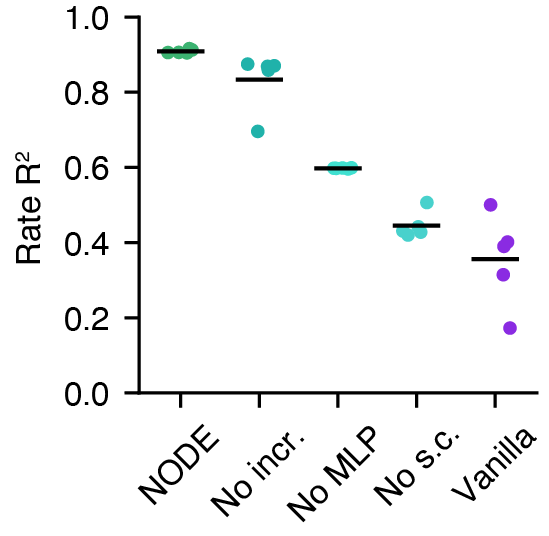}
  \caption{Performance of discretized and ablated NODE-based models on the Arneodo dataset. Ablated versions of our model were trained on the Arneodo dataset (Fig. \ref{fig:arneodo1}) from five different initializations. Learning rates were tuned for each model class, dropout was fixed to 0.05, and weight decay was off. NODE denotes a discretized version of our model. "No incr." ablated incremental training, "No MLP" ablated the two-layer MLP, and "No s.c." ablated skip connections. "Vanilla" models used vanilla RNNs, equivalent to ablating all of the above.}
  \label{fig:ablation}
\end{figure}

To illustrate the roles that incremental training, skip connections, and the MLP play in the NODE's performance, we trained ablated models on the Arneodo dataset (Fig. \ref{fig:ablation}). We show only rate $R^2$ for these models, because models must fit the data well before state $R^2$ becomes relevant. Removing the skip connection and replacing the MLP component with the forward pass of a vanilla RNN had independent and similarly severe effects on model performance. Note that without the MLP, the discretized NODE has significant similarity to a GRU with appropriately biased gates ($\mathbf{r}_t=1$ and $\mathbf{z}_t=1-\Delta=0.9$). Removing incremental training had mild effects on performance.

\section{Discussion}
Many high-dimensional artificial and biological neural networks (NNs) have been shown to solve tasks using relatively simple, low-dimensional dynamics. Thus, a natural hypothesis is that when artificial NNs are trained to reproduce the activity of biological NNs, they will recover the same low-dimensional dynamics. Our results show that this is not necessarily the case-- the architecture of the artificial NN plays an important role in parsimonious and accurate recovery. In this work we demonstrated that widely-used, RNN-based architectures struggle to accurately recover the latent, low-dimensional dynamical systems underlying simulated neural population activity. In contrast, the high-capacity MLPs and skip connections used by NODE-based architectures enable them to recover the latent dynamics where RNNs fail.

On one hand, RNNs are an intuitive abstraction for an interacting population of biological neurons. They explicitly represent the activation of and relationships between units over time, and they can be optimized to test hypotheses about how computational dynamics can be used to solve various tasks. This insight has enabled a substantial body of interpretability work in recent years. However, task-based modeling is different from interpretable neural population modeling in several important ways that might explain why RNNs seem to perform well with the former but struggle with the latter. First, the output targets of task-based models are often lower-dimensional and dynamically simpler than that of neural population models (NPMs). The former often use on a handful of fixed or step-like inputs and outputs, while the latter rely on high-dimensional activity with complex temporal patterns. Generally, the task-based paradigm should impose fewer constraints on the the RNNs, making them easier to optimize and regularize. Second, the requirements of the dynamics module of an interpretable NPM are more demanding than that of a task-based model. For interpretable NPMs, the goal is to learn a representation of latent dynamics with high parsimony and accuracy, which we have shown is difficult for RNNs.

By testing our models on multiple datasets, we also gain insight into how dataset properties can affect dynamical recovery. First, we found that global dynamics may be underdetermined given a limited set of locally sampled latent trajectories, resulting in incomplete recovery of fixed points distant from the sampled area (e.g., for the R\"{o}ssler system, Fig. \ref{fig:rossler_lorenz}b). Second, for some datasets, models tended to achieve noticeably better performance with larger latent spaces (e.g., for the Lorenz system, Fig. \ref{fig:rossler_lorenz}d, top). This could potentially bias researchers to overestimate the dimensionality of the latent dynamics, and result in inaccurate recovery. We suspect this issue may relate to the level of chaos in a given system. For the systems tested, the different maximum Lyapunov exponents, which characterize the rate of divergence of infinitesimally close initial conditions, support this idea. For R\"{o}ssler, Arneodo, and Lorenz, they are  0.076, 0.243, and 1.091 respectively \cite{rossler_equation_1976, arneodo_occurence_1980, lorenz_deterministic_1963, gilpin_chaos_2021}. We hypothesize that extra model dimensions may provide performance gains in the presence of chaos by allowing correction of divergent trajectories caused by small initial condition placement errors. More generally, we caution that while NODE-based SAEs reduce the risk involved in model selection, the models with the best reconstruction performance are not guaranteed to most concisely capture the true dynamics. This problem may be an unavoidable consequence of the increased flexibility that comes with higher latent dimensionality.

Most of the NODE-based SAEs trained in this work used a continuous-time forward pass. From a practical perspective, computing the forward pass with a solver required more evaluations of the vector field and slowed down training by about 4x compared to the GRU-based SAEs (15 mins vs. 1 hour for 3k epochs). However, we found that the continuous-time forward pass is not critical for performance (Fig. \ref{fig:ablation}), and that the NODE-based SAE can be easily discretized at the bin width with an appropriate step size to close this gap. Models appear to be equally sensitive to learning rate and have comparable compute requirements and both seem to benefit slightly from incremental training. While the number of parameters in the GRU-based SAE is fixed, the NODE-based SAE allows the modeler to dial in the parameters and capacity by specifying the depth and width of the MLP.

The models used here differ from state-of-the-art RNN-based NPMs like latent factor analysis via dynamical systems (LFADS) \cite{pandarinath_inferring_2018} in several key ways. At a high level, LFADS also incorporates inferred inputs, variational inference, and a low-dimensional bottleneck in $g$. From an interpretability perspective, inference of inputs is a difficult problem due to essential ambiguities about whether a given change in state is driven by inputs or internal dynamics. While LFADS made some progress by imposing reasonable priors on inferred inputs, more work is necessary to ensure that inputs are inferred with high accuracy. Related, the most salient feature of variational inference for LFADS is that information flow through the network can be controlled via KL penalties on initial conditions and inferred inputs. This enables the contributions of internal dynamics and inputs to be balanced via hyperparameter tuning. Thus, variational inference may play a more important role in the context of inferred inputs. The low-dimensional bottleneck used in LFADS allows the dimensionality of a larger generator RNN to be decoupled from the latent dimensionality of the inferred rates. In our model, a bottleneck is not necessary because the dynamics themselves can be learned in a low-dimensional space. However, both the LFADS readout and the NODE-based SAE readout may be inadequate for capturing the complexity of real neural manifolds. Using a linear readout on neural data that lies on a nonlinear manifold could impose additional burden on the latent dynamics model, potentially requiring higher dimensionality or higher complexity. While some of the differences between our NODE-based SAE and LFADS could provide a path toward improved performance and applicability, care should be taken to preserve interpretability with each extension of the model.

Despite the demonstrated success of our method on synthetic data, a few considerations remain regarding its applicability to real neural data. First, unlike our simulated systems, real neural populations are constantly receiving input from other brain areas, meaning that we can't take for granted that their dynamics are time-invariant. A second but related question is whether our techniques for interpretation, particularly FP linearization, are  appropriate tools for understanding the brain.

To the first point, the key to revealing computational dynamics \textit{in vivo} will likely be careful experimental design that monitors specific brain areas during specific tasks, designed to isolate the hypothesized dynamics to the monitored population. Perhaps further experiments on synthetic data could demonstrate that recovery is achievable for partially observed and interacting dynamical systems.

With respect to the second point, recent experiments that optogenetically perturb neural activity have found that FPs are not just convenient abstractions, but that they capture an underlying dynamical structure that is predictive of the behavior of the neural circuit \cite{inagaki_discrete_2019, finkelstein_attractor_2021}. Thus, direct inspection of vector fields and FP analyses restricted to regions that are well populated by nearby data are still promising approaches to understanding neural dynamics in real data.

\section*{Acknowledgments}
The authors would like to acknowledge Timothy D. Kim and Carlos Brody for helpful discussions that further developed the ideas in this manuscript.

\clearpage

\bibliography{main.bib}

\clearpage

\begin{center}
  \Large\textbf{Supplementary Material}
\end{center}

\section{Datasets}

\subsection{Dynamical systems}

\subsubsection{Arneodo}

\begin{align}
    \dot{x} &= y \\
    \dot{y} &= z \\
    \dot{z} &= -a x - b y - c z +d x^3
\end{align}

where $a=-5.5$, $b=4.5$, $c=1.0$, and $d=-1.0$ \cite{arneodo_occurence_1980}.

\subsubsection{R\"{o}ssler}

\begin{align}
    \dot{x} &= -y - z \\
    \dot{y} &= x + ay \\
    \dot{z} &= b + z(x - c)
\end{align}

where $a=0.2$, $b=0.2$, and $c=5.7$ \cite{rossler_equation_1976}.

\subsubsection{Lorenz}

\begin{align}
    \dot{x} &= \sigma (y-x) \\
    \dot{y} &= x (\rho -z)-y \\
    \dot{z} &= x y-\beta z
\end{align}

where $\beta=2.667$, $\rho=28$, and $\sigma=10$ \cite{lorenz_deterministic_1963}.

\subsubsection{Mackey-Glass}

\begin{equation}
    \dot{x} = \beta\frac{x_\tau}{1 + x_\tau^n} - \gamma x
\end{equation}

where $x_\tau=x(t-\tau)$, $\beta=2$, $\gamma=1.0$, $n=9.65$, and $\tau=2.0$ \cite{glass1979pathological}. Note that the system defined by this delay differential equation was converted to $D=10$ by stacking past time steps: $\mathbf{x}(t)=[x(t), x(t-\tau)... x(t-(D-1)\tau)]$.

\subsubsection{Mac Arthur}

The Mac Arthur system models a set of species competing for renewing resources \cite{macarthur1969species}. We use parameters taken from a model of phytoplankton \cite{huisman1999biodiversity}, where $n=5$ distinct populations $N_i$ compete for access to $k=5$ distinct resource types $R_j$.

\begin{align}
    \dot{N}_i &= N_i * (\mu_i(R_1,\dots,R_k) - m_i) \quad i=1,\dots,n \\
    \dot{R}_j &= d * (\mathbf{s}_j - R_j) - \sum_{i=1}^n \mathbf{C}_{ji}\mu_i(R_1,\dots,R_k)N_i \quad j=1,\dots,k \\
\end{align}

where

\begin{align}
    \mu_i(R_1,\dots,R_k) &= \min\left(\frac{r_i R_1}{\mathbf{K}_{1i} + R_1},\dots,\frac{r_i R_k}{\mathbf{K}_{ki} + R_k},\right) \\
    \mathbf{C} &= 
    \begin{bmatrix}
        0.04 & 0.04 & 0.07 & 0.04 & 0.04 \\
        0.08 & 0.08 & 0.08 & 0.10 & 0.08 \\
        0.10 & 0.10 & 0.10 & 0.10 & 0.14 \\
        0.05 & 0.03 & 0.03 & 0.03 & 0.03 \\
        0.07 & 0.09 & 0.07 & 0.07 & 0.07 \\
    \end{bmatrix} \\
    \mathbf{K} &= 
    \begin{bmatrix}
        0.39 & 0.34 & 0.30 & 0.24 & 0.23 \\
        0.22 & 0.39 & 0.34 & 0.30 & 0.27 \\
        0.27 & 0.22 & 0.39 & 0.34 & 0.30 \\
        0.30 & 0.24 & 0.22 & 0.39 & 0.34 \\
        0.34 & 0.30 & 0.22 & 0.20 & 0.39 \\
    \end{bmatrix} \\
    \mathbf{s} &= 
    \begin{bmatrix}
        0.04 & 0.04 & 0.07 & 0.04 & 0.04 \\
    \end{bmatrix} \\
    m_i &= 0.25 \\
    d &= 0.25 \\
    r_i &= 1 \\
\end{align}

We take the state of the system to be the 10-D vector $\mathbf{x} = [N_1, \dots, N_n, R_1, \dots, R_k]$. Unlike other systems, the Mac Arthur trajectories had dramatically different means and variances across dimensions. To remedy this, we standardized each dimension to zero mean and unit variance before projecting into the neural dimension.

\subsection{Solving and resampling latent trajectories}

Systems were simulated using the \texttt{dysts} Python package, which offered well-reasoned standards for initial conditions, integration steps, and resampling frequency \cite{gilpin_chaos_2021}. Initial conditions had been determined by running each model until the moments of the autocorrelation function were stationary. Integration steps had been chosen based on the highest significant frequency observed in the power spectrum (see Table \ref{table:1}). After integration, trajectories were resampled to contain a specified number of points per period (see Table \ref{table:1}), where period was defined based on the dominant frequency in each system's power spectrum.

\begin{table}[h!]
\caption{Dataset parameters and characteristics}
\centering
\begin{tabular}{|c||c c c c c|} 
 \hline
 System & Arneodo & R\"{o}ssler & Lorenz & Mackey-Glass & Mac Arthur \\
 \hline
 Solver step size (t) & 1.215e-3 & 7.563e-4 & 1.801e-4 & 1.156e-3 & 4.384e-3 \\ 
 Points per period & 35 &  35 & 70 & 35 & 70 \\
 Mean firing rate (spikes/bin) & \makecell{1.624 ($N=10$) \\ 1.641 ($N=100$)} & 1.872 & 1.695 & 1.609 & 1.606 \\
 \hline
\end{tabular}
\label{table:1}
\end{table}

\section{Models}

\subsection{Gated Recurrent Unit} \label{supp:gru}

Each GRU cell computes the following:

\begin{align}
    r_t &= \sigma(W_{ir} x_t + b_{ir} + W_{hr} h_{(t-1)} + b_{hr}) \\
    z_t &= \sigma(W_{iz} x_t + b_{iz} + W_{hz} h_{(t-1)} + b_{hz}) \\
    n_t &= \tanh(W_{in} x_t + b_{in} + r_t * (W_{hn} h_{(t-1)}+ b_{hn})) \\
    h_t &= (1 - z_t) * n_t + z_t * h_{(t-1)}
\end{align}

where $h_t$ is the hidden state, $h_{(t-1)}$ is the previous hidden state, $x_t$ is the input, $r_t$ is the reset gate, $z_t$ is the update gate, $n_t$ is the candidate state, and $\sigma$ is the sigmoid function.

\section{Training details}

All weights were initialized from $\mathcal{U}(-\sqrt{k}, \sqrt{k})$, where $k=1/\mathrm{in\_features}$ for linear layers and $k=1/\mathrm{hidden\_size}$ for GRU weights. Dropout layers ($p=0.05$) were inserted before and after the IC linear projection during training. Loss was computed as the average Poisson NLL across neurons and time points and models were trained by gradient descent using Adam on batches of 650 samples for 3000 epochs. Further training details for all models are given in Table \ref{supp:hp_table}.

\begin{table}[h!]
\caption{Training hyperparameters}
\centering
\begin{tabular}{|c||c c|c|c c c c c|} 
 \hline
  & \multicolumn{2}{c|}{Main Experiments} & Mac Arthur & & & Ablation & & \\
 \hline
  & GRU & NODE & NODE & NODE & No incr. & No MLP & No s.c. & Vanilla  \\
 \hline
 Learning Rate & 1e-3 & 5e-3 & 5e-4 & 5e-4 & 1e-2 & 1e-2 & 1e-3 & 1e-2 \\
 Weight Decay  & 1e-4 & 1e-5 & N/A & N/A & N/A & N/A & N/A & N/A \\
 Points per group & N/A & 5 & 20 & 20 & N/A & 20  & 20 & N/A \\
 Epochs per group & N/A & 150 & 500 & 500 & N/A & 500  & 500 & N/A \\
\hline
\end{tabular}
\label{supp:hp_table}
\end{table}

\section{Finding fixed points}

For each network, fixed points were located by randomly sampling 1024 initial states from inferred latent trajectories and iteratively minimizing the squared norm of the vector field \cite{sussillo_opening_2013}. We used Adam with a learning rate of 1e-2 for minimization over 5000 iterations. Uniqueness was determined using a distance threshold of 1e-3 and points that did not achieve a squared norm less than 1e-10 were excluded. In practice, there were many orders of magnitude between kept and excluded points.

\section{Compute resources}

We used an internal computing cluster with a total of 30 NVIDIA GeForce RTX 2080 Ti GPUs for model training. Continuous-time NODE-based models took approximately 45 minutes to train and discretized NODE-based and GRU-based models took approximately 15 minutes. Our experiments trained 175 of the former and 270 of the latter (excluding hyperparameter tuning). With 3 models training on each GPU, running these experiments took approximately 65 GPU hours.

\section{Metrics}

\subsection{Rate reconstruction}

We computed the coefficient of determination between true ($\mathbf{Y}$) and predicted ($\hat{\mathbf{Y}}$) rates for each neuron, and reported the average value across neurons. We refer to this metric as rate $R^2$.

\[
    \text{Rate} \: R^2 = R^2(\mathbf{Y}, \hat{\mathbf{Y}}) = \frac{1}{N}\sum_{n=1}^{N} 1 -  \frac{\sum_{i=1}^{T}(y_{i,n} - \hat{y}_{i,n})^2}{\sum_{i=1}^{T}(y_{i,n} - \bar{y}_n)^2}
\]

\subsection{State reconstruction}

To compute state $R^2$, we concatenated a column of ones to the true latent states ($\mathbf{Z}_1$), then used the pseudoinverse to find a set of weights that produced an affine transform from the true latent states to the inferred latent states ($\hat{\mathbf{Z}}$) with minimal squared error. We computed $R^2$ for this projection as described in the previous section.

\begin{align}
    \mathbf{W}_z &= \mathbf{Z}_1^\dagger \hat{\mathbf{Z}} \\
    \text{State} \: R^2 &= R^2(\hat{\mathbf{Z}}, \mathbf{Z}_1 \mathbf{W}_z)
\end{align}

\section{Open-source packages used}

\begin{itemize}
  \item \href{https://github.com/pytorch/pytorch}{\texttt{torch}} \cite{paszke_pytorch_2019} (BSD license): Deep learning framework providing layer definitions, GPU acceleration, automatic differentiation, optimization, and more.
  \item \href{https://github.com/PyTorchLightning/pytorch-lightning}{\texttt{pytorch\_lightning}} (Apache 2.0 license): Lightweight wrappers for model training.
  \item \href{https://github.com/ray-project/ray}{\texttt{ray.tune}} \cite{liaw2018tune} (Apache 2.0 license): Distributed hyperparameter tuning.
  \item \href{https://github.com/williamgilpin/dysts}{\texttt{dysts}} \cite{gilpin_chaos_2021} (Apache 2.0 license): Implementations for modeled dynamical systems.
  \item \href{https://github.com/DiffEqML/torchdyn}{\texttt{torchdyn}} \cite{poli2020torchdyn} (Apache 2.0 license): Implementation of NODEs.
  \item \href{https://github.com/mattgolub/fixed-point-finder}{\texttt{fixed\_point\_finder}} \cite{Golub2018} (Apache 2.0 license): Inspiration for \texttt{torch}-based fixed point finder.
\end{itemize}

\end{document}